\documentclass[%
 aip,
 amsmath,amssymb,
 reprint,
]{revtex4-1}
\pdfoutput=1

\usepackage{sfmath}
\usepackage{graphicx}
\usepackage{dcolumn}
\usepackage{bm}
\usepackage{dblfloatfix}
\usepackage[dvipsnames]{xcolor}
\newcommand\myshade{85}
\definecolor{a}{HTML}{05bc61}
\usepackage{hyperref}
\hypersetup{
    linkcolor  = a!\myshade!black,
  citecolor  = a!\myshade!black,
  urlcolor   = a!\myshade!black,
  colorlinks = true,
}
\newcommand{\angstrom}{\text{\normalfont\AA}}
\usepackage[utf8]{inputenc}
\usepackage{amsmath}
\usepackage{setspace}
\usepackage{siunitx}
\usepackage[T1]{fontenc}

\begin{document}

\title{Mapping Electron Beam-Induced Radiolytic Damage in Molecular Crystals}
\author{Ambarneil Saha}
\email{asaha2@lbl.gov
}
\affiliation{National Center for Electron Microscopy, Molecular Foundry, Lawrence Berkeley National Laboratory, Berkeley, California 94720, United States}

\author{Matthew Mecklenburg}
\affiliation{California NanoSystems Institute, University of California, Los Angeles, California 90095, United States}

\author{Alexander J. Pattison}
\affiliation{National Center for Electron Microscopy, Molecular Foundry, Lawrence Berkeley National Laboratory, Berkeley, California 94720, United States}

\author{Aaron S. Brewster}
\affiliation{Molecular Biophysics and Integrated Bioimaging Division, Lawrence Berkeley National Laboratory, Berkeley, California 94720, United States}

\author{Jos\'e A. Rodr\'iguez}
\email{jrodriguez@mbi.ucla.edu}
\affiliation{Department of Chemistry and Biochemistry, University of California, Los Angeles, California 90095, United States}
\affiliation{UCLA--DOE Institute for Genomics and Proteomics, University of California, Los Angeles, California 90095, United States}

\author{Peter Ercius}
\email{percius@lbl.gov}
\affiliation{National Center for Electron Microscopy, Molecular Foundry, Lawrence Berkeley National Laboratory, Berkeley, California 94720, United States}

\date{April 26, 2024}

\begin{abstract}
Every electron crystallography experiment is fundamentally limited by radiation damage. Nevertheless, little is known about the onset and progression of radiolysis in beam-sensitive molecular crystals. Here we apply ambient-temperature scanning nanobeam electron diffraction to record simultaneous dual-space snapshots of organic and organometallic nanocrystals at sequential stages of beam-induced radiolytic decay. We show that the underlying mosaic of coherently diffracting zones (CDZs) continuously undergoes spatial reorientation as a function of accumulating electron exposure, causing the intensities of many Bragg reflections to fade nonmonotonically. Furthermore, we demonstrate that repeated irradiation at a single probe position leads to the concentric propagation of delocalized radiolytic damage well beyond the initial point of impact. These results sharpen our understanding of molecular crystals as conglomerates of CDZs whose complex lattice structure deteriorates through a series of dynamic internal changes during exposure to ionizing radiation.
 
\end{abstract}

\maketitle

\section{Introduction} 
\begin{figure*}
\includegraphics[width=7.1in]{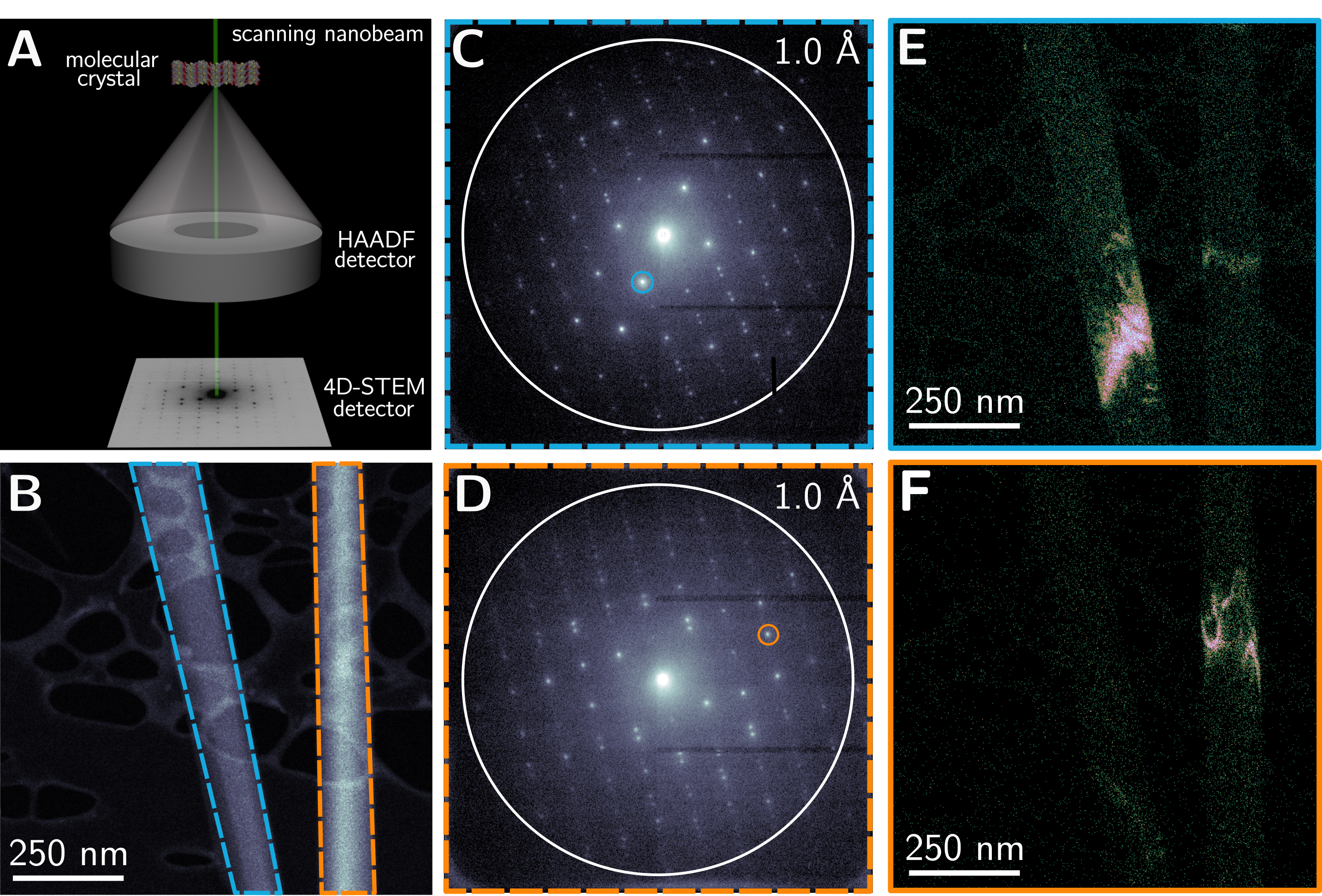}
\caption{\textsf{\label{fig1} \textbf{Overview of scanning nanobeam electron diffraction.} (A) Schematic of a 4D--STEM experiment. (B) HAADF--STEM image depicting two nanocrystals of biotin. The dashed lines superimposed on each crystal represent the boundaries of two real-space virtual apertures. (C, D) Summed diffraction patterns exported from the regions outlined in (B) blue and orange, respectively. The blue and orange circles enclosing Bragg reflections represent reciprocal-space virtual apertures. (E, F) Virtual dark-field images reconstructed from the Bragg reflections encircled in (C) blue and (D) orange, spatially mapping which coherently diffracting zones produced those peaks.}}
\end{figure*}

\noindent Immediately as a molecular crystal is illuminated within a transmission electron microscope, it undergoes sustained bombardment with extremely hazardous levels of ionizing $\beta$-radiation.\cite{stenn_specimen_1970} Electrons accelerated to relativistic speeds typically deposit quantities of energy per unit mass in the range of megagrays (MGy; $10^{6}$ J kg$^{-1}$) \cite{Henderson_1995}, several orders of magnitude greater than the lethal Gy-scale doses (10 J kg$^{-1}$) associated with nuclear disasters.\cite{drozdovitch_radiation_2021} Thus, electron beam-induced radiolytic damage has long been considered \cite{baker_chapter_2010} the “fundamental limit” constraining fields such as single-particle analysis, \cite{cheng_primer_2015} cryo-electron tomography, \cite{turk_promise_2020} and electron crystallography. \cite{saha_electron_2022}\\
\indent Despite its crucial impact, little is rigorously understood about the onset and progression of radiolytic damage in beam-sensitive materials, particularly in real space. If a primary inelastic collision causes electronic excitation of an atom within the specimen, many deleterious consequences can ensue.\cite{egerton_radiation_2019} These include direct ionization, excitation and dissipation of collective plasmon and phonon oscillations, secondary electron emission, and homolytic scission of covalent bonds.\cite{grubb_radiation_1974, flannigan_pulsed-beam_2023} Arguably the most consequential downstream effect of primary excitation is the generation, diffusion, and propagation of itinerant free radicals, generally thought to occur on the picosecond timescale.\cite{grubb_radiation_1974, flannigan_pulsed-beam_2023} Once formed, these reactive species can diffuse throughout the specimen and spark cascades of secondary homolytic reactions. As a result, this process is often considered the predominant mechanism by which radiolytic damage propagates in molecular crystals.\cite{talmon_electron_1987, fryer_effect_1987, meents_h2}\\
\indent In electron crystallography, the primary metric used for assessing the extent of degradation induced by radiolysis is the disappearance of Bragg reflections in electron diffraction patterns.\cite{glaeser_limitations_1971} Upon prolonged exposure, the intensities of Bragg peaks recede into the noise floor, indicating the permanent loss of periodicity in the lattice. Several previous studies have analyzed the radiolytic decay of Bragg reflections in organic and biomolecular crystals.\cite{glaeser_limitations_1971, reimer_interpretation_1982, li_radiation_2004, hattne_analysis_2018, sari_toward_2018, peet_energy_2019, naydenova_reduction_2022} Critically, however, the scope of these analyses has mostly been limited to indirect observation via the back focal plane, and little is known about the concomitant changes in real space which \textit{drive} the deterioration of Bragg peaks. Since accurate measurement of Bragg peak intensities is a key prerequisite to \textit{ab initio} structure determination,\cite{khouchen_optimal_2023} a deeper understanding of radiolytic decay in molecular crystals is necessary. Nevertheless, a \textit{simultaneous} visualization of the effects of electron beam-induced radiolysis—in both real space and reciprocal space—remains elusive.\\
\indent Here we leverage scanning nanobeam electron diffraction—also known as 4D scanning transmission electron microscopy (4D--STEM)—to provide a dual-space perspective on this longstanding problem. 4D--STEM involves raster-scanning a focused electron probe across a user-selected region of interest in real space, partitioned into an $n \times n$ grid of scan points.\cite{bustillo_4D-STEM_2021, ophus_four-dimensional_2019} At each probe position in this two-dimensional grid, an independent diffraction pattern is recorded on a backthinned CMOS-based direct electron detector operating at a frame rate of 87 kHz, corresponding to a readout speed of 11  $\mu$s.\cite{ercius_4d_2020} Simultaneously, a conventional monolithic high-angle annular dark-field (HAADF) detector is used to acquire a thickness map (Figure \ref{fig1}A). This approach yields a four-dimensional dataset consisting of $n^{2}$ sparse diffraction patterns (indexed by reciprocal-space coordinates $q_{x}$ and $q_{y}$), each linked with an associated pair of localized probe positions ($x, y$) in real space. By computationally summing signal from user-selected regions of a crystal (Figure \ref{fig1}B), 4D--STEM empowers us to construct custom virtual apertures (Figure \ref{fig1}C-D),\cite{gammer_diffraction_2015} functionally permitting direct visualization of exactly which coherently diffracting zones (CDZs) gave rise to Bragg signal in reciprocal space (Figure \ref{fig1}E-F).\\
\indent Furthermore, our experiments exploit recent improvements in detector sensitivity and speed\cite{ercius_4d_2020} to add a critical fifth dimension: time. Rapid acquisition and live analysis of large volumes of data (e.g., up to 13200 GB) is facilitated by a unique file-transfer and data reduction pipeline involving sparsification of the raw detector frames on supercomputer nodes. This procedure---amounting to nearly lossless compression---produces lightweight, MB-scale files (\href{https://zenodo.org/records/10387146}{Zenodo repository}) directly viewable in our interactive graphical user interface, \href{https://github.com/ercius/DuSC_explorer}{\texttt{DuSC\_explorer}}. By recording a series of consecutive 4D--STEM scans on the same specimen, here we explicitly visualize that the CDZs comprising molecular crystals undergo internal spatial rearrangement as a function of accumulating exposure to electron beam-induced radiolysis.

\section{Disparate patterns of decay}
\begin{figure*}
\hspace{-0.7cm}
\includegraphics[width=7.1in]{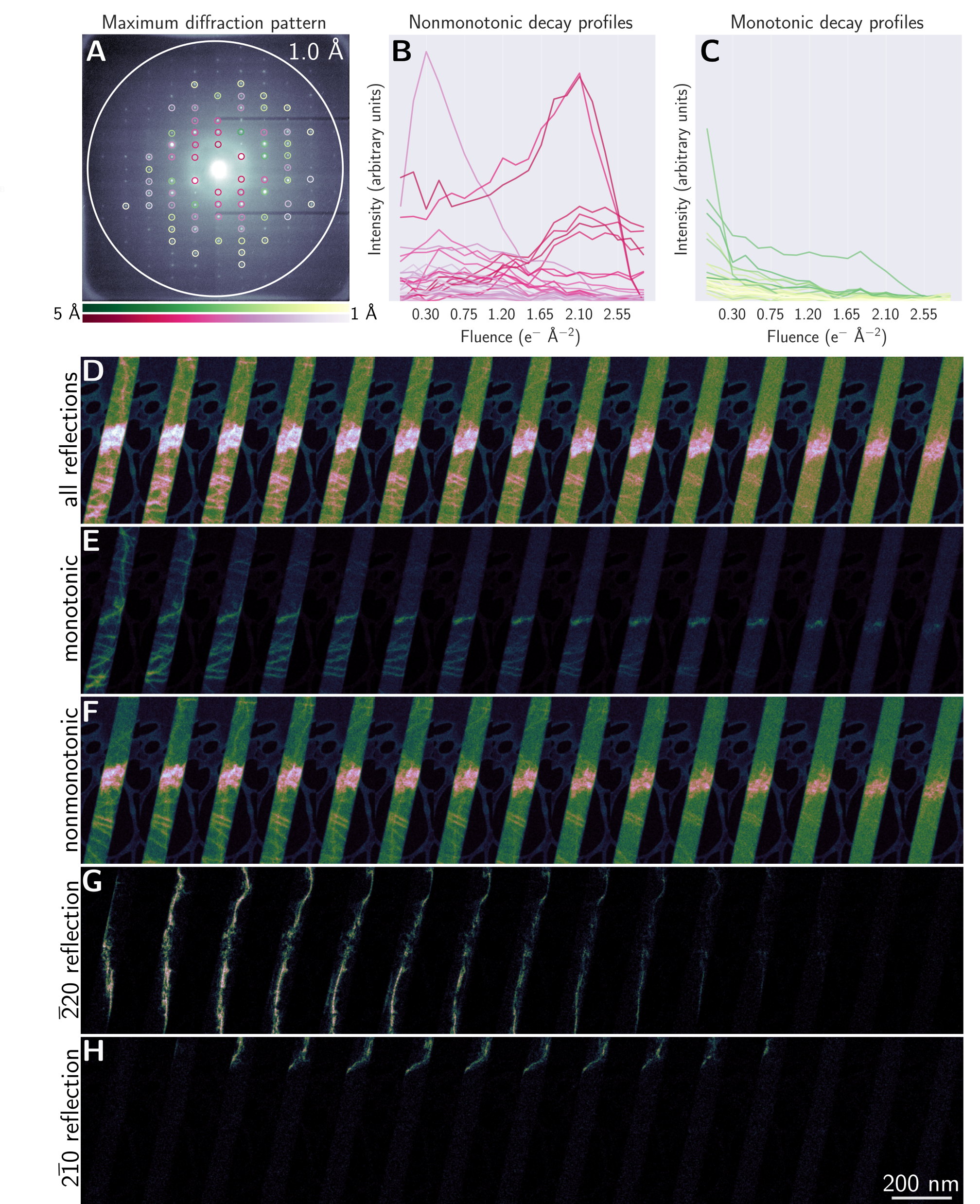}
\caption{\textsf{\label{fig2} \textbf{Rearrangement of coherently diffracting zones in biotin.} (A) Maximum diffraction pattern across all scans in the time series. (B, C) Decay profiles of Bragg reflections undergoing (B) nonmonotonic decay and (C) monotonic decay, corresponding to the peaks encircled in (B) red and (C) green in (A). (D--H) Horizontal montages of VDF images reconstructed by placing reciprocal-space virtual apertures around (D) all 68 Bragg reflections identified in (A), (E) 38 Bragg reflections undergoing nonmonotonic decay, (F) 30 Bragg reflections undergoing monotonic decay, (G) a single Bragg reflection undergoing nonmonotonic decay which is present in the first scan and transiently grows stronger, and (H) a single Bragg reflection undergoing nonmonotonic decay which first appears in the third scan. Each VDF image represents a consecutive 4D-STEM experiment.}}
\end{figure*}

\noindent We began our studies by conducting 25 consecutive low-fluence 4D--STEM experiments (0.15 $e^{-} \angstrom^{-2}$ scan$^{-1}$) at RT on nanocrystals of biotin (Table S1), a beam-sensitive organic molecule (Figure S1) typically used as a standard in 3D electron crystallography.\cite{bruhn2023} As long-range order in the biotin lattice was destroyed, we observed that new diffraction signal was frequently unmasked from dormant, initially Bragg-silent CDZs. We found that beam-induced spatial reorientation causes a continuously shifting array of reciprocal lattice vectors to intersect the surface of the Ewald sphere, leading to several distinct patterns of decay. Broadly, we classify these into two groups: monotonic and nonmonotonic decay. To identify these two groups of reflections, we generated and indexed (Figure S2) a synthetic diffraction pattern where the intensity of each pixel represented a projection of its maximum value during the time series (Figure \ref{fig2}A). We then placed virtual apertures around the centroids of 68 clearly identifiable Bragg reflections and tracked their intensities as a function of fluence (Figure \ref{fig2}B-C). Next, we reconstructed virtual dark-field (VDF) images to visualize the corresponding real-space signal, displayed as a horizontal montage with increasing electron exposure applied from left to right (Figure \ref{fig2}D-F).\\
\indent Intriguingly, in the composite VDF images corresponding to all 68 reflections, we observed an intricate network of veins and striations meandering across the biotin crystal, with an especially intense and long-lived swath located at the center (Figure \ref{fig2}D). When summing signal exclusively from the 30 reflections obeying monotonic decay, however, the intense central band dimmed considerably (Figure \ref{fig2}E), unveiling a weaker group of striated CDZs underneath. Since reflections undergoing monotonic decay fade earlier than their nonmonotonic counterparts (Figure \ref{fig2}B-C), we reasoned that the most robust CDZs likely produce Bragg peaks inclined to experience nonmonotonic decay. As expected, the VDF images corresponding to the CDZs associated with the 38 nonmonotonically deteriorating reflections recapitulated most of the real-space signal in the central band (Figure \ref{fig2}F). Furthermore, a clear resolution-dependent pattern emerged: stronger, lower-resolution reflections tended to decay nonmonotonically, whereas weaker, higher-resolution reflections tended to decay monotonically.\\
\indent After reconstructing the aggregate signal from clusters of reflections, we then moved on to case studies of individual Bragg peaks. In monotonic decay, the recorded intensity for a given Bragg peak has already reached its global maximum in the first 4D--STEM scan. From this apex, it then undergoes simple, nearly linear deterioration. For instance, the twisted striation in Movie S1—corresponding to the [$\bar{2}20$] Bragg reflection ($d$ = 2.36 Å)—migrates to the right fringe of the crystal while thinning rapidly. As the intensity of this Bragg peak becomes indistinguishable from noise, its CDZ disintegrates in lockstep, indicating that reflections which undergo monotonic decay originate from CDZs which also diminish monotonically in size (Movie S1).\\
\indent Conversely, in nonmonotonic decay, the global maximum of the recorded Bragg peak intensity is linearly offset from the first scan, resulting in a delayed-onset decay pattern which often resembles a skewed Gaussian distribution. These reflections represent reciprocal lattice vectors which initially do not precisely fulfill the Bragg condition. Nevertheless, their excitation error is effectively reduced in whatever radiolysis-induced orientation the lattice has progressed to in a later scan, causing them to paradoxically appear stronger for some ephemeral period before ultimately succumbing to monotonic decay. For instance, the Friedel mate of the 2.36 Å peak considered earlier—i.e., reflection [$2\bar{2}0$]—follows a nonmonotonic decay pattern (Figure \ref{fig2}G). Although this CDZ is morphologically similar to its Friedel mate, it unfurls into a more expansive striation in scans 2 and 3 relative to its shape in scan 1. Consistent with this transient boost in CDZ size, reflection [$2 \bar{2}0$] also endures an extra 0.45 $e^{-} \angstrom^{-2}$ compared to its Friedel mate before fading (Movie S2).\\
\indent Another subset of nonmonotonic decay involves Bragg peaks absent in scan 1 which proceed to materialize midway through the dose series before deteriorating. For instance, the crescent-shaped CDZ in Figure \ref{fig2}H—corresponding to reflection [$2\bar{1}0$]—is nonexistent in scan 1 and begins to form only after approximately 0.3 $e^{-} \angstrom^{-2}$ have been delivered (Movie S3). This CDZ then extends across the surface of the crystal and appears to recede back into the upper-left corner as it decays. Although this 2.61 Å reflection’s Friedel mate [$\bar{2}10$] activates a similar nonmonotonic decay pattern, its lifetime is a truncated subset of its counterpart in Figure \ref{fig2}H, and its shape is not as expansive (Movie S4). This disparate behavior mirrors the unexpected asymmetry observed in the decay patterns of reflections [$2\bar{2}0$] and [$\bar{2}20$]. At no point did an individual Bragg reflection uniformly activate the entirety of the biotin crystal, indicating clear lattice heterogeneity\cite{boggon_synchrotron_2000} even at the nanoscale.\cite{gallagher-jones_nanoscale_2019} Collectively, our analysis indicates that the nanoscale topography of the molecular lattice is constantly shifting immediately following the onset of radiolysis. 

\section{Delocalized propagation of damage}
\begin{figure*}
\hspace{-0.7cm}
\includegraphics[width=7.1in]{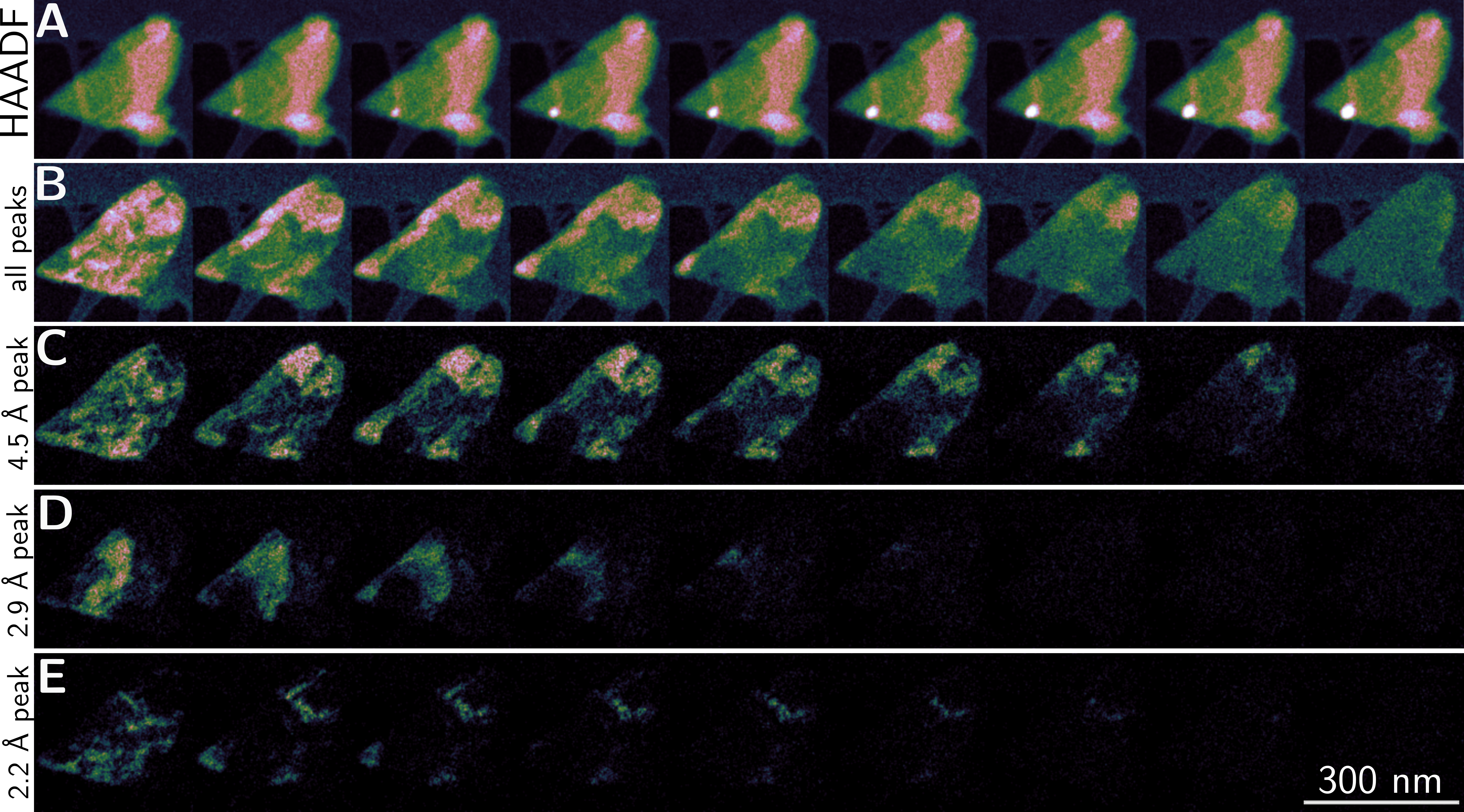}
\caption{\textsf{\label{fig3} \textbf{Delocalized propagation of radiolytic damage in Ni(dppf)Cl$_{2}$.} (A) Horizontal montage of HAADF--STEM images simultaneously acquired alongside 4D--STEM data. (B--E) Horizontal montages of VDF images reconstructed by placing reciprocal-space virtual apertures around (B) all clearly identifiable Bragg reflections, (C) a single 4.5 Å reflection, (D) a single 2.9 Å reflection, and (E) a single 2.2 Å reflection.}}
\end{figure*}

\noindent In our initial 4D--STEM scans on biotin, each probe position received an equivalent fluence, similar to the uniform illumination delivered in conventional parallel-beam TEM. We then sought to explore whether 4D--STEM could visualize the radiolytic damage caused by asymmetric delivery of incident electrons. For these experiments, we amplified the incident fluence (1.5 $e^{-} \angstrom^{-2}$ scan$^{-1}$) and switched substrates to a more beam-stable organometallic complex\cite{jones_organomet} based on a ferrocene scaffold, Ni(dppf)Cl$_{2}$ (Figure S1, Table S2). Our first experiment involved using an approximately 18 nm (full-width at half maximum) focused STEM probe to irradiate one single region of the Ni(dppf)Cl$_{2}$ crystal for an extra 200 ms between the acquisition of 300 consecutive 4D–STEM datasets. This created a highly localized “dead zone” in which diffraction signal was quickly extinguished, despite the simultaneous presence of strong Bragg reflections originating from active CDZs distal to this region.\\
\indent Concurrently acquired HAADF--STEM images (Figure \ref{fig3}A) showed no evidence of mass loss, instead revealing a small buildup of carbon contamination which proceeded to increase in intensity throughout all subsequent scans. Intriguingly, the corresponding VDF images reconstructed from all Bragg peaks displayed a tide of amorphization radially propagating from this area, suggesting that the “dead zone” at the dwell point acts as a concentrated reservoir of reactive free radicals (Figure \ref{fig3}B). With each consecutive scan, these destructive radicals diffused into areas of the lattice increasingly far removed from the original source, further ablating all CDZs in their path. Each time, their advances were halted by molecular bulwarks such as aromatic rings, which can temporarily arrest a cascade of further radiolytic reactions via several postulated quenching pathways.\cite{alexander_energy_1954, liming_hydrogen-addition_1968, lin_electron_1974, howie_electron_1985} Nevertheless, since these mechanisms cannot quench propagating free radicals or secondary electrons indefinitely, each consecutive 4D–STEM scan maps the outward progression of an increasingly delocalized frontier of amorphization.\\
\indent Crucially, we then discovered that the extent of delocalization from the dwell point also varies as a function of scattering angle. VDF images reconstructed from a series of progressively higher-resolution Bragg reflections—at 4.45 Å (Figure \ref{fig3}C, Movie S5), 2.94 Å (Figure \ref{fig3}D, Movie S6), and 2.22 Å (Figure \ref{fig3}E, Movie S7), respectively (Figure S3)—unveiled that the dimensions of the growing “impact crater” were clearly larger in the latter two cases for an identical point in the time series. CDZs producing low-resolution Bragg peaks generally appear larger in size and stronger in diffracting power than their high-resolution counterparts, mostly because they correspond to a more crude level of long-range order present more uniformly throughout the lattice. Consequently, these CDZs tend to dominate the real-space signal in composite VDF images generated from all reflections. Therefore, this resolution-dependent effect is visible only when placing virtual apertures around individual peaks and is obscured when summing aggregate signal from groups of reflections.\\
\indent Our physical interpretation of this phenomenon is that a small vanguard of itinerant free radicals diffuses significantly farther than the frontier delineated by the composite VDF images—on the order of tens to hundreds of nm—before being quenched. Although these species represent a minority of overall radicals produced, the weaker CDZs producing high-resolution reflections require fewer radiolytic reactions to destroy, since these CDZs correspond to a granular level of short-range order distributed much more sporadically within the lattice. As a result, the progress of the tide of amorphization is resolution-dependent, and the radius of the “impact crater” expands when reconstructing VDF images from higher-resolution Bragg reflections. Using a higher fluence, we also successfully reproduced the delocalized propagation of amorphization in two related Pd-based organometallic complexes, Pd(dcypf)Cl$_{2}$ (Figure S4) and Pd(dppf)Cl$_{2}$ (Figure S5).\\
\indent An intriguing counterpoint to the delocalization visualized in this work is provided by macromolecular X-ray crystallography. Mechanistic nuances aside, both electrons and X-rays ultimately inflict damage to molecular crystals via radiolysis.\cite{Henderson_1995} Nevertheless, radiolytic damage caused by X-rays is often confined to the volume directly irradiated by the impinging quanta.\cite{Schulze-Briese:xn0007} We attribute this discrepancy to the severe mismatch in size between the $\mu$m-scale FWHM of the incident X-ray beam and the nm-scale propagation of radiolytic damage observed in our RT scanning nanobeam electron diffraction experiments. In sum, this phenomenon is only visible when (a) the incident beam consistently delivers more fluence to one specific region of the illuminated crystal and (b) the dimensions of the probe roughly match the mean free path of whatever actively propagating reactive species is generated via radiolysis---likely dominated by free radicals at RT.

\section{Implications for 3D ED}
\noindent These observations lead to critical implications for conventional 3D electron diffraction (3D ED) experiments, where an irradiated nanocrystal is continuously rotated while reciprocal space is periodically sampled.\cite{saha_electron_2022} We hypothesized that a judicious amount of beam-induced radiolysis could potentially anneal the initial distribution of CDZs into a more self-consistent lattice, transiently producing higher-quality data. To test this, we collected a series of 3D ED datasets on two species we had also surveyed by 4D–STEM: Ni(dppf)Cl$_{2}$ (Figure S6) and biotin (Figure S7). In Ni(dppf)Cl$_{2}$, we found that three key metrics typically used to assess crystallographic data quality—$R_{\text{meas}}$, $<$I/$\sigma$(I)$>$, and CC$_{1/2}$—improved for the second pass relative to the first pass, followed by a steady decline. As a robust measurement of internal consistency,\cite{diederichs_improved_1997} lower $R_{\text{meas}}$ is particularly suggestive of the possibility that radiolysis can attenuate the signal from rotationally misoriented CDZs, leading to an ensemble of more accurately measured intensities produced from a more monolithic lattice. Conversely, recording Bragg peak intensities while CDZ rearrangement is actively occurring—i.e., while the underlying distribution of CDZs is still continuously morphing due to beam-induced radiolysis—is analogous to taking a blurry photograph of a moving object. This causes CC$_{1/2}$ and $R_{\text{meas}}$ to suffer, rationalizing the discrepancy between the first and second passes.\\
\indent In other words, if the substrate under interrogation is sufficiently beam-stable, a full 3D ED movie is obtainable during the critical window after CDZ reorientation has stabilized but before monotonic decay has attenuated the sub-1.2 Å resolution reflections necessary for \textit{ab initio} structure solution. Ni(dppf)Cl$_{2}$ nicely fits these criteria. In especially beam-sensitive samples, however, this lag phase\cite{owen_exploiting_2014} is likely exceptionally narrow. For instance, even slow-cooled to 100 K, CDZ rearrangement in biotin proved too quick to productively exploit, as most CDZs had already progressed deep into the monotonic decay phase by the time a second-pass movie could be acquired.

\section{Spectroscopic corroboration}
\noindent Recognizing that diffraction alone cannot provide a complete picture of the chemical effects of radiolytic damage,\cite{li_radiation_2004} we then set out to briefly investigate whether electron energy-loss spectroscopy (EELS) could serve as a parallel metric for assessing beam-induced chemical destruction in these systems. In particular, aromaticity is known to produce an EELS resonance at an energy loss of approximately 6 eV, corresponding to the same $\pi$--$\pi^{*}$ electronic transition typically probed by ultraviolet-visible spectroscopy.\\
\indent We used parallel-beam TEM to record both EEL spectra and selected-area electron diffraction (SAED) patterns from the same crystal of dppf (Figure S8). After the initial acquisition of an EEL spectrum from an individual nanocrystal, each specimen was continuously irradiated until the corresponding SAED patterns had completely faded. Subsequent acquisition of another EEL spectrum (Figure S9) revealed that the 6 eV $\pi$--$\pi^{*}$ peak remained intact, confirming that crystallographic destruction precedes chemical destruction in these species.\\
\indent Furthermore, our EEL spectra indicate that a typical inelastic collision during an electron diffraction experiment involves an energy loss of approximately 22 eV (Figure S10). To contextualize, this value is 4--5$\times$ higher than the homolytic bond dissociation enthalpy of an average sp$^{3}$ C—F bond (4--5 eV), generally considered one of the strongest bonds in organic chemistry.\cite{ohagan_understanding_2008} Although it is unclear whether this progression of events will hold in nonaromatic compounds, we note that the recent development of monochromated STEM beams with sub-10 meV resolution in the infrared regime\cite{hachtel_identification_2019} provides an excellent range of spectroscopic handles (e.g., the decay of a C—H stretching mode) for assessing this in greater chemical detail.

\section{Discussion}
\noindent Several previous analyses of electron beam-induced radiolytic damage in organic crystals have coalesced around models involving monotonic decay of Bragg peak intensities.\cite{li_radiation_2004, peet_energy_2019, naydenova_reduction_2022} Nevertheless, most literature precedent exclusively monitors the fading of Bragg reflections, with little to no corroborating analysis conducted in real space. Our 4D–STEM results expand this body of work by directly visualizing a crucial missing piece in this puzzle: beam-induced spatial reorientation, driven by damaged CDZs engaging in a complex dance choreographed by radiolysis. A key consequence of this effect is that monotonic decay—the ultimate fate of every reflection—is initially coincident with an ephemeral phase where radiolytic damage reshuffles the intensities of some Bragg peaks. Subsequently, after some rotationally misoriented CDZs have been amorphized, the remainder of the lattice exhibits a more stagnant orientation, and the intensities of the associated Bragg reflections fade monotonically.\\
\indent In 1971, Glaeser disclosed an analogous scenario in uranyl acetate-stained catalase crystals, where he observed “relative changes in diffraction intensities” upon irradiation by 80 keV electrons.\cite{glaeser_limitations_1971} Glaeser’s hypothesis that a “significant portion of the matter within the unit cell of the structure changes to some more stable configuration” is partially consistent with our 4D–STEM data. Slight changes within individual unit cells—caused, for instance, by a global increase in atomic $B$-factors, or by site-specific damage to particular functional groups—would affect the intensities of the recorded Bragg peaks. Nevertheless, unit cell-level effects cannot adequately rationalize the sudden appearance of Bragg reflections absent in earlier scans (Figure \ref{fig2}H), and \textit{ab initio} structures determined from multipass 3D ED datasets do not display drastic deviations in the RMSDs of atomic positions.\cite{hattne_analysis_2018} Furthermore, evidence from EEL spectra (Figure S8--10) indicates some preservation of chemical identity even after total loss of crystallinity. Therefore, our results point more definitively to an internal rearrangement of the CDZs themselves—i.e., at the level of lattice subregions or \textit{ensembles} of unit cells—as the driving force behind nonmonotonic decay. Similar precedent for nonmonotonic fluctuations in Bragg peak intensities has also been documented in the context of macromolecular X-ray crystallography.\cite{owen_exploiting_2014, warkentin_global_2013, warkentin_lifetimes_2017} \\
\indent Interestingly, Chiu and coworkers have reported that transient strengthening of Fourier amplitudes in frozen-hydrated catalase is more widespread at 4 K than 100 K.\cite{bammes_radiation_2010} Chiu and coworkers categorized these effects as “abnormal” and proposed specimen movement or charging as possible explanations. Our 4D--STEM analysis points to electron beam-induced CDZ reorientation as an equally plausible interpretation of their results. In all our datasets, a salient finding illustrated by the simultaneously acquired HAADF--STEM images is that the nanoscale morphology of the specimens stayed completely intact, with no significant variations in shape, size, thickness, or spatial position. These results indicate that only internal changes within the lattice provide an explanation fully consistent with the experimental data.\\ 
\indent Moreover, our experiments suggest that the nonmonotonic decay of Bragg reflections is a straightforward consequence of the migration of CDZs caused by radiolytic damage. Although a detailed investigation of temperature is outside the scope of our current work, we speculate that some fraction of weaker Bragg reflections which appear to deteriorate monotonically at room temperature genuinely undergo some type of nonmonotonic decay. In these cases, the progression of radiolytic damage is simply too quick, and the corresponding CDZs have already shifted before the first 4D--STEM scan is complete. Nevertheless, Chiu and coworkers’ observation of a greater proportion of nonmonotonically fading Fourier amplitudes at lower temperatures suggests that slower propagation of free radicals may lead to an elongation of the period in which CDZ rearrangement is visible.\\

\section{Conclusion}
\noindent In a landmark 1970 review, Stenn and Bahr lamented the seemingly intractable difficulties posed by radiolytic damage within the transmission electron microscope.\cite{stenn_specimen_1970} These researchers opined that “it is near impossible to follow or predict the secondary events occurring in the electron microscopic specimen because of [the] intensely complex cascading process” involved, concluding that “we shall have to be satisfied with knowing relatively little.” Although the underlying phenomena remain as complex as ever, the advent of rapid,\cite{ercius_4d_2020} low-fluence 4D-STEM has enabled us to directly visualize the onset and progression of radiolytic damage in molecular nanocrystals, all in simultaneous dual-space detail. By unveiling the decay dynamics of coherently diffracting zones associated with nonmonotonically fading Bragg reflections, these findings reshape our current understanding of radiolytic damage, paving the way for novel strategies aimed at mitigating these effects. Furthermore, our discovery of expanding “impact craters” at the nanoscale provides a powerful technique to visualize the delocalized spread of amorphization, building a platform for detailed mechanistic studies of radiolysis enabled by 4D--STEM.

\begin{acknowledgments}

\noindent The authors thank S. Welborn, B. Enders, C. Harris, and S. Nia for technical support, as well as R. Egerton, R. Glaeser, J. Dickerson, and E. Garman for useful discussions.\\
\indent \textit{Funding}: Work at the Molecular Foundry was supported by the Office of Science, Office of Basic Energy Sciences, of the U.S. Department of Energy under Contract No. DE-AC02-05CH11231. This research used resources of the National Energy Research Scientific Computing Center (NERSC), a U.S. Department of Energy Office of Science user facility located at Lawrence Berkeley National Laboratory, operated under Contract No. DE-AC02-05CH11231. This work was supported by the Laboratory Directed Research and Development Program of Lawrence Berkeley National Laboratory under U.S. Department of Energy Contract No. DE-AC02-05CH11231, the BioPACIFIC Materials Innovation Platform of the National Science Foundation under Award No. DMR-1933487, and the National Science Foundation Graduate Research Fellowship (to A.S.) under Grant No. DGE-2034835. Work by J.A.R. was supported by a Packard Fellowship, U.S. Department of Energy Grant No. DE-FC02-02ER63421, NIH-NIGMS Grant No. R35 GM128867, and the HHMI Emerging Pathogens Initiative.
\\
\indent \textit{Competing interests}: J.A.R. is a founder and equity stakeholder of MEDSTRUC INC.\\
\indent \textit{Data availability}: All 4D--STEM data is publicly accessible on Zenodo (DOI: \href{https://zenodo.org/records/10387146}{\texttt{10.5281/zenodo.10387146}}) and viewable using \href{https://github.com/ercius/DuSC_explorer}{\texttt{DuSC\_explorer}}, our open-source python-based software.\\
\indent \textit{CRediT author statement}: 
\begin{itemize}
\item{Conceptualization: A.S., J.A.R., and P.E.} 
\item{Methodology: A.S., J.A.R., and P.E.} 
\item{Formal analysis: A.S., M.M., A.S.B., J.A.R., and P.E.}
\item{Investigation: A.S., M.M., A.J.P., and P.E.} 
\item{Data curation: A.S. and P.E.}
\item{Software: A.S., A.J.P., and P.E.}
\item{Writing—original draft and preparation: A.S.}
\item{Writing—review and editing: A.S., with input from all authors.} \item{Visualization: A.S. and P.E.}
\item{Supervision: J.A.R. and P.E.}
\item{Funding acquisition: J.A.R. and P.E.}
\end{itemize}

\end{acknowledgments}
\newpage

\bibliography{main}

\providecommand{\noopsort}[1]{}\providecommand{\singleletter}[1]{#1}%
\begin{thebibliography}{41}%
\makeatletter
\providecommand \@ifxundefined [1]{%
 \@ifx{#1\undefined}
}%
\providecommand \@ifnum [1]{%
 \ifnum #1\expandafter \@firstoftwo
 \else \expandafter \@secondoftwo
 \fi
}%
\providecommand \@ifx [1]{%
 \ifx #1\expandafter \@firstoftwo
 \else \expandafter \@secondoftwo
 \fi
}%
\providecommand \natexlab [1]{#1}%
\providecommand \enquote  [1]{``#1''}%
\providecommand \bibnamefont  [1]{#1}%
\providecommand \bibfnamefont [1]{#1}%
\providecommand \citenamefont [1]{#1}%
\providecommand \href@noop [0]{\@secondoftwo}%
\providecommand \href [0]{\begingroup \@sanitize@url \@href}%
\providecommand \@href[1]{\@@startlink{#1}\@@href}%
\providecommand \@@href[1]{\endgroup#1\@@endlink}%
\providecommand \@sanitize@url [0]{\catcode `\\12\catcode `\$12\catcode
  `\&12\catcode `\#12\catcode `\^12\catcode `\_12\catcode `\%12\relax}%
\providecommand \@@startlink[1]{}%
\providecommand \@@endlink[0]{}%
\providecommand \url  [0]{\begingroup\@sanitize@url \@url }%
\providecommand \@url [1]{\endgroup\@href {#1}{\urlprefix }}%
\providecommand \urlprefix  [0]{URL }%
\providecommand \Eprint [0]{\href }%
\providecommand \doibase [0]{http://dx.doi.org/}%
\providecommand \selectlanguage [0]{\@gobble}%
\providecommand \bibinfo  [0]{\@secondoftwo}%
\providecommand \bibfield  [0]{\@secondoftwo}%
\providecommand \translation [1]{[#1]}%
\providecommand \BibitemOpen [0]{}%
\providecommand \bibitemStop [0]{}%
\providecommand \bibitemNoStop [0]{.\EOS\space}%
\providecommand \EOS [0]{\spacefactor3000\relax}%
\providecommand \BibitemShut  [1]{\csname bibitem#1\endcsname}%
\let\auto@bib@innerbib\@empty
\bibitem [{\citenamefont {Stenn}\ and\ \citenamefont
  {Bahr}(1970)}]{stenn_specimen_1970}%
  \BibitemOpen
  \bibfield  {author} {\bibinfo {author} {\bibfnamefont {K.}~\bibnamefont
  {Stenn}}\ and\ \bibinfo {author} {\bibfnamefont {G.~F.}\ \bibnamefont
  {Bahr}},\ }\bibfield  {title} {\enquote {\bibinfo {title} {Specimen damage
  caused by the beam of the transmission electron microscope, a correlative
  reconsideration},}\ }\href {\doibase 10.1016/S0022-5320(70)90167-X}
  {\bibfield  {journal} {\bibinfo  {journal} {\textit{Journal of Ultrastructure
  Research}}\ }\textbf {\bibinfo {volume} {31}},\ \bibinfo {pages} {526--550}
  (\bibinfo {year} {1970})}\BibitemShut {NoStop}%
\bibitem [{\citenamefont {Henderson}(1995)}]{Henderson_1995}%
  \BibitemOpen
  \bibfield  {author} {\bibinfo {author} {\bibfnamefont {R.}~\bibnamefont
  {Henderson}},\ }\bibfield  {title} {\enquote {\bibinfo {title} {The potential
  and limitations of neutrons, electrons and x-rays for atomic resolution
  microscopy of unstained biological molecules},}\ }\href {\doibase
  10.1017/S003358350000305X} {\bibfield  {journal} {\bibinfo  {journal}
  {\textit{Quarterly Reviews of Biophysics}}\ }\textbf {\bibinfo {volume}
  {28}},\ \bibinfo {pages} {171–193} (\bibinfo {year} {1995})}\BibitemShut
  {NoStop}%
\bibitem [{\citenamefont {Drozdovitch}(2021)}]{drozdovitch_radiation_2021}%
  \BibitemOpen
  \bibfield  {author} {\bibinfo {author} {\bibfnamefont {V.}~\bibnamefont
  {Drozdovitch}},\ }\bibfield  {title} {\enquote {\bibinfo {title} {Radiation
  exposure to the thyroid after the {Chernobyl} accident},}\ }\href
  {https://www.frontiersin.org/articles/10.3389/fendo.2020.569041} {\bibfield
  {journal} {\bibinfo  {journal} {\textit{Frontiers in Endocrinology}}\
  }\textbf {\bibinfo {volume} {11}} (\bibinfo {year} {2021})}\BibitemShut
  {NoStop}%
\bibitem [{\citenamefont {Baker}\ and\ \citenamefont
  {Rubinstein}(2010)}]{baker_chapter_2010}%
  \BibitemOpen
  \bibfield  {author} {\bibinfo {author} {\bibfnamefont {L.~A.}\ \bibnamefont
  {Baker}}\ and\ \bibinfo {author} {\bibfnamefont {J.~L.}\ \bibnamefont
  {Rubinstein}},\ }\bibfield  {title} {\enquote {\bibinfo {title} {{Chapter
  Fifteen--Radiation Damage in Electron Cryomicroscopy}},}\ }in\ \href
  {\doibase 10.1016/S0076-6879(10)81015-8} {\emph {\bibinfo {booktitle}
  {Methods in Enzymology}}},\ \bibinfo {series} {Cryo-{EM} Part A Sample
  Preparation and Data Collection}, Vol.\ \bibinfo {volume} {481},\ \bibinfo
  {editor} {edited by\ \bibinfo {editor} {\bibfnamefont {G.~J.}\ \bibnamefont
  {Jensen}}}\ (\bibinfo  {publisher} {Academic Press},\ \bibinfo {year}
  {2010})\ pp.\ \bibinfo {pages} {371--388}\BibitemShut {NoStop}%
\bibitem [{\citenamefont {Cheng}\ \emph {et~al.}(2015)\citenamefont {Cheng},
  \citenamefont {Grigorieff}, \citenamefont {Penczek},\ and\ \citenamefont
  {Walz}}]{cheng_primer_2015}%
  \BibitemOpen
  \bibfield  {author} {\bibinfo {author} {\bibfnamefont {Y.}~\bibnamefont
  {Cheng}}, \bibinfo {author} {\bibfnamefont {N.}~\bibnamefont {Grigorieff}},
  \bibinfo {author} {\bibfnamefont {P.~A.}\ \bibnamefont {Penczek}}, \ and\
  \bibinfo {author} {\bibfnamefont {T.}~\bibnamefont {Walz}},\ }\bibfield
  {title} {\enquote {\bibinfo {title} {A primer to single-particle
  cryo-electron microscopy},}\ }\href {\doibase 10.1016/j.cell.2015.03.050}
  {\bibfield  {journal} {\bibinfo  {journal} {\textit{Cell}}\ }\textbf
  {\bibinfo {volume} {161}},\ \bibinfo {pages} {438--449} (\bibinfo {year}
  {2015})}\BibitemShut {NoStop}%
\bibitem [{\citenamefont {Turk}\ and\ \citenamefont
  {Baumeister}(2020)}]{turk_promise_2020}%
  \BibitemOpen
  \bibfield  {author} {\bibinfo {author} {\bibfnamefont {M.}~\bibnamefont
  {Turk}}\ and\ \bibinfo {author} {\bibfnamefont {W.}~\bibnamefont
  {Baumeister}},\ }\bibfield  {title} {\enquote {\bibinfo {title} {The promise
  and the challenges of cryo-electron tomography},}\ }\href {\doibase
  10.1002/1873-3468.13948} {\bibfield  {journal} {\bibinfo  {journal}
  {\textit{{FEBS} Letters}}\ }\textbf {\bibinfo {volume} {594}},\ \bibinfo
  {pages} {3243--3261} (\bibinfo {year} {2020})}\BibitemShut {NoStop}%
\bibitem [{\citenamefont {Saha}, \citenamefont {Nia},\ and\ \citenamefont
  {Rodríguez}(2022)}]{saha_electron_2022}%
  \BibitemOpen
  \bibfield  {author} {\bibinfo {author} {\bibfnamefont {A.}~\bibnamefont
  {Saha}}, \bibinfo {author} {\bibfnamefont {S.~S.}\ \bibnamefont {Nia}}, \
  and\ \bibinfo {author} {\bibfnamefont {J.~A.}\ \bibnamefont {Rodríguez}},\
  }\bibfield  {title} {\enquote {\bibinfo {title} {Electron diffraction of {3D}
  molecular crystals},}\ }\href {\doibase 10.1021/acs.chemrev.1c00879}
  {\bibfield  {journal} {\bibinfo  {journal} {\textit{Chemical Reviews}}\
  }\textbf {\bibinfo {volume} {122}},\ \bibinfo {pages} {13883--13914}
  (\bibinfo {year} {2022})}\BibitemShut {NoStop}%
\bibitem [{\citenamefont {Egerton}(2019)}]{egerton_radiation_2019}%
  \BibitemOpen
  \bibfield  {author} {\bibinfo {author} {\bibfnamefont {R.~F.}\ \bibnamefont
  {Egerton}},\ }\bibfield  {title} {\enquote {\bibinfo {title} {Radiation
  damage to organic and inorganic specimens in the {TEM}},}\ }\href {\doibase
  10.1016/j.micron.2019.01.005} {\bibfield  {journal} {\bibinfo  {journal}
  {\textit{Micron}}\ }\textbf {\bibinfo {volume} {119}},\ \bibinfo {pages}
  {72--87} (\bibinfo {year} {2019})}\BibitemShut {NoStop}%
\bibitem [{\citenamefont {Grubb}(1974)}]{grubb_radiation_1974}%
  \BibitemOpen
  \bibfield  {author} {\bibinfo {author} {\bibfnamefont {D.~T.}\ \bibnamefont
  {Grubb}},\ }\bibfield  {title} {\enquote {\bibinfo {title} {Radiation damage
  and electron microscopy of organic polymers},}\ }\href {\doibase
  10.1007/BF00540772} {\bibfield  {journal} {\bibinfo  {journal}
  {\textit{Journal of Materials Science}}\ }\textbf {\bibinfo {volume} {9}},\
  \bibinfo {pages} {1715--1736} (\bibinfo {year} {1974})}\BibitemShut {NoStop}%
\bibitem [{\citenamefont {Flannigan}\ and\ \citenamefont
  {{VandenBussche}}(2023)}]{flannigan_pulsed-beam_2023}%
  \BibitemOpen
  \bibfield  {author} {\bibinfo {author} {\bibfnamefont {D.~J.}\ \bibnamefont
  {Flannigan}}\ and\ \bibinfo {author} {\bibfnamefont {E.~J.}\ \bibnamefont
  {{VandenBussche}}},\ }\bibfield  {title} {\enquote {\bibinfo {title}
  {Pulsed-beam transmission electron microscopy and radiation damage},}\ }\href
  {\doibase 10.1016/j.micron.2023.103501} {\bibfield  {journal} {\bibinfo
  {journal} {\textit{Micron}}\ }\textbf {\bibinfo {volume} {172}},\ \bibinfo
  {pages} {103501} (\bibinfo {year} {2023})}\BibitemShut {NoStop}%
\bibitem [{\citenamefont {Talmon}(1987)}]{talmon_electron_1987}%
  \BibitemOpen
  \bibfield  {author} {\bibinfo {author} {\bibfnamefont {Y.}~\bibnamefont
  {Talmon}},\ }\bibfield  {title} {\enquote {\bibinfo {title} {Electron beam
  radiation damage to organic and biological cryospecimens},}\ }in\ \href
  {\doibase 10.1007/978-3-642-72815-0_3} {\emph {\bibinfo {booktitle}
  {Cryotechniques in Biological Electron Microscopy}}},\ \bibinfo {editor}
  {edited by\ \bibinfo {editor} {\bibfnamefont {R.~A.}\ \bibnamefont
  {Steinbrecht}}\ and\ \bibinfo {editor} {\bibfnamefont {K.}~\bibnamefont
  {Zierold}}}\ (\bibinfo  {publisher} {Springer},\ \bibinfo {year} {1987})\
  pp.\ \bibinfo {pages} {64--84}\BibitemShut {NoStop}%
\bibitem [{\citenamefont {Fryer}(1987)}]{fryer_effect_1987}%
  \BibitemOpen
  \bibfield  {author} {\bibinfo {author} {\bibfnamefont {J.~R.}\ \bibnamefont
  {Fryer}},\ }\bibfield  {title} {\enquote {\bibinfo {title} {The effect of
  dose rate on imaging aromatic organic crystals},}\ }\href {\doibase
  10.1016/0304-3991(87)90242-7} {\bibfield  {journal} {\bibinfo  {journal}
  {\textit{Ultramicroscopy}}\ }\textbf {\bibinfo {volume} {23}},\ \bibinfo
  {pages} {321--327} (\bibinfo {year} {1987})}\BibitemShut {NoStop}%
\bibitem [{\citenamefont {Meents}\ \emph {et~al.}(2010)\citenamefont {Meents},
  \citenamefont {Gutmann}, \citenamefont {Wagner},\ and\ \citenamefont
  {Schulze-Briese}}]{meents_h2}%
  \BibitemOpen
  \bibfield  {author} {\bibinfo {author} {\bibfnamefont {A.}~\bibnamefont
  {Meents}}, \bibinfo {author} {\bibfnamefont {S.}~\bibnamefont {Gutmann}},
  \bibinfo {author} {\bibfnamefont {A.}~\bibnamefont {Wagner}}, \ and\ \bibinfo
  {author} {\bibfnamefont {C.}~\bibnamefont {Schulze-Briese}},\ }\bibfield
  {title} {\enquote {\bibinfo {title} {Origin and temperature dependence of
  radiation damage in biological samples at cryogenic temperatures},}\ }\href
  {\doibase 10.1073/pnas.0905481107} {\bibfield  {journal} {\bibinfo  {journal}
  {\textit{Proceedings of the National Academy of Sciences}}\ }\textbf
  {\bibinfo {volume} {107}},\ \bibinfo {pages} {1094--1099} (\bibinfo {year}
  {2010})}\BibitemShut {NoStop}%
\bibitem [{\citenamefont {Glaeser}(1971)}]{glaeser_limitations_1971}%
  \BibitemOpen
  \bibfield  {author} {\bibinfo {author} {\bibfnamefont {R.~M.}\ \bibnamefont
  {Glaeser}},\ }\bibfield  {title} {\enquote {\bibinfo {title} {Limitations to
  significant information in biological electron microscopy as a result of
  radiation damage},}\ }\href {\doibase 10.1016/S0022-5320(71)80118-1}
  {\bibfield  {journal} {\bibinfo  {journal} {\textit{Journal of Ultrastructure
  Research}}\ }\textbf {\bibinfo {volume} {36}},\ \bibinfo {pages} {466--482}
  (\bibinfo {year} {1971})}\BibitemShut {NoStop}%
\bibitem [{\citenamefont {Reimer}\ and\ \citenamefont
  {Spruth}(1982)}]{reimer_interpretation_1982}%
  \BibitemOpen
  \bibfield  {author} {\bibinfo {author} {\bibfnamefont {L.}~\bibnamefont
  {Reimer}}\ and\ \bibinfo {author} {\bibfnamefont {J.}~\bibnamefont
  {Spruth}},\ }\bibfield  {title} {\enquote {\bibinfo {title} {Interpretation
  of the fading of diffraction patterns from organic substances irradiated with
  100 {keV} electrons at 10-300 {K}},}\ }\href {\doibase
  10.1016/0304-3991(82)90039-0} {\bibfield  {journal} {\bibinfo  {journal}
  {\textit{Ultramicroscopy}}\ }\textbf {\bibinfo {volume} {10}},\ \bibinfo
  {pages} {199--210} (\bibinfo {year} {1982})}\BibitemShut {NoStop}%
\bibitem [{\citenamefont {Li}\ and\ \citenamefont
  {Egerton}(2004)}]{li_radiation_2004}%
  \BibitemOpen
  \bibfield  {author} {\bibinfo {author} {\bibfnamefont {P.}~\bibnamefont
  {Li}}\ and\ \bibinfo {author} {\bibfnamefont {R.~F.}\ \bibnamefont
  {Egerton}},\ }\bibfield  {title} {\enquote {\bibinfo {title} {Radiation
  damage in coronene, rubrene and p-terphenyl, measured for incident electrons
  of kinetic energy between 100 and 200 ke{V}},}\ }\href {\doibase
  10.1016/j.ultramic.2004.05.010} {\bibfield  {journal} {\bibinfo  {journal}
  {\textit{Ultramicroscopy}}\ }\textbf {\bibinfo {volume} {101}},\ \bibinfo
  {pages} {161--172} (\bibinfo {year} {2004})}\BibitemShut {NoStop}%
\bibitem [{\citenamefont {Hattne}\ \emph {et~al.}(2018)\citenamefont {Hattne},
  \citenamefont {Shi}, \citenamefont {Glynn}, \citenamefont {Zee},
  \citenamefont {Gallagher-Jones}, \citenamefont {Martynowycz}, \citenamefont
  {Rodriguez},\ and\ \citenamefont {Gonen}}]{hattne_analysis_2018}%
  \BibitemOpen
  \bibfield  {author} {\bibinfo {author} {\bibfnamefont {J.}~\bibnamefont
  {Hattne}}, \bibinfo {author} {\bibfnamefont {D.}~\bibnamefont {Shi}},
  \bibinfo {author} {\bibfnamefont {C.}~\bibnamefont {Glynn}}, \bibinfo
  {author} {\bibfnamefont {C.-T.}\ \bibnamefont {Zee}}, \bibinfo {author}
  {\bibfnamefont {M.}~\bibnamefont {Gallagher-Jones}}, \bibinfo {author}
  {\bibfnamefont {M.~W.}\ \bibnamefont {Martynowycz}}, \bibinfo {author}
  {\bibfnamefont {J.~A.}\ \bibnamefont {Rodriguez}}, \ and\ \bibinfo {author}
  {\bibfnamefont {T.}~\bibnamefont {Gonen}},\ }\bibfield  {title} {\enquote
  {\bibinfo {title} {Analysis of global and site-specific radiation damage in
  cryo-{EM}},}\ }\href {\doibase 10.1016/j.str.2018.03.021} {\bibfield
  {journal} {\bibinfo  {journal} {\textit{Structure}}\ }\textbf {\bibinfo
  {volume} {26}},\ \bibinfo {pages} {759--766} (\bibinfo {year}
  {2018})}\BibitemShut {NoStop}%
\bibitem [{\citenamefont {S’ari}\ \emph {et~al.}(2018)\citenamefont
  {S’ari}, \citenamefont {Blade}, \citenamefont {Brydson}, \citenamefont
  {Cosgrove}, \citenamefont {Hondow}, \citenamefont {Hughes},\ and\
  \citenamefont {Brown}}]{sari_toward_2018}%
  \BibitemOpen
  \bibfield  {author} {\bibinfo {author} {\bibfnamefont {M.}~\bibnamefont
  {S’ari}}, \bibinfo {author} {\bibfnamefont {H.}~\bibnamefont {Blade}},
  \bibinfo {author} {\bibfnamefont {R.}~\bibnamefont {Brydson}}, \bibinfo
  {author} {\bibfnamefont {S.~D.}\ \bibnamefont {Cosgrove}}, \bibinfo {author}
  {\bibfnamefont {N.}~\bibnamefont {Hondow}}, \bibinfo {author} {\bibfnamefont
  {L.~P.}\ \bibnamefont {Hughes}}, \ and\ \bibinfo {author} {\bibfnamefont
  {A.}~\bibnamefont {Brown}},\ }\bibfield  {title} {\enquote {\bibinfo {title}
  {Toward developing a predictive approach to assess electron beam instability
  during transmission electron microscopy of drug molecules},}\ }\href
  {\doibase 10.1021/acs.molpharmaceut.8b00693} {\bibfield  {journal} {\bibinfo
  {journal} {\textit{Molecular Pharmaceutics}}\ }\textbf {\bibinfo {volume}
  {15}},\ \bibinfo {pages} {5114--5123} (\bibinfo {year} {2018})}\BibitemShut
  {NoStop}%
\bibitem [{\citenamefont {Peet}, \citenamefont {Henderson},\ and\ \citenamefont
  {Russo}(2019)}]{peet_energy_2019}%
  \BibitemOpen
  \bibfield  {author} {\bibinfo {author} {\bibfnamefont {M.~J.}\ \bibnamefont
  {Peet}}, \bibinfo {author} {\bibfnamefont {R.}~\bibnamefont {Henderson}}, \
  and\ \bibinfo {author} {\bibfnamefont {C.~J.}\ \bibnamefont {Russo}},\
  }\bibfield  {title} {\enquote {\bibinfo {title} {The energy dependence of
  contrast and damage in electron cryomicroscopy of biological molecules},}\
  }\href {\doibase 10.1016/j.ultramic.2019.02.007} {\bibfield  {journal}
  {\bibinfo  {journal} {\textit{Ultramicroscopy}}\ }\textbf {\bibinfo {volume}
  {203}},\ \bibinfo {pages} {125--131} (\bibinfo {year} {2019})}\BibitemShut
  {NoStop}%
\bibitem [{\citenamefont {Naydenova}\ \emph {et~al.}(2022)\citenamefont
  {Naydenova}, \citenamefont {Kamegawa}, \citenamefont {Peet}, \citenamefont
  {Henderson}, \citenamefont {Fujiyoshi},\ and\ \citenamefont
  {Russo}}]{naydenova_reduction_2022}%
  \BibitemOpen
  \bibfield  {author} {\bibinfo {author} {\bibfnamefont {K.}~\bibnamefont
  {Naydenova}}, \bibinfo {author} {\bibfnamefont {A.}~\bibnamefont {Kamegawa}},
  \bibinfo {author} {\bibfnamefont {M.~J.}\ \bibnamefont {Peet}}, \bibinfo
  {author} {\bibfnamefont {R.}~\bibnamefont {Henderson}}, \bibinfo {author}
  {\bibfnamefont {Y.}~\bibnamefont {Fujiyoshi}}, \ and\ \bibinfo {author}
  {\bibfnamefont {C.~J.}\ \bibnamefont {Russo}},\ }\bibfield  {title} {\enquote
  {\bibinfo {title} {On the reduction in the effects of radiation damage to
  two-dimensional crystals of organic and biological molecules at liquid-helium
  temperature},}\ }\href {\doibase 10.1016/j.ultramic.2022.113512} {\bibfield
  {journal} {\bibinfo  {journal} {\textit{Ultramicroscopy}}\ }\textbf {\bibinfo
  {volume} {237}},\ \bibinfo {pages} {113512} (\bibinfo {year}
  {2022})}\BibitemShut {NoStop}%
\bibitem [{\citenamefont {Khouchen}\ \emph {et~al.}(2023)\citenamefont
  {Khouchen}, \citenamefont {Klar}, \citenamefont {Chintakindi}, \citenamefont
  {Suresh},\ and\ \citenamefont {Palatinus}}]{khouchen_optimal_2023}%
  \BibitemOpen
  \bibfield  {author} {\bibinfo {author} {\bibfnamefont {M.}~\bibnamefont
  {Khouchen}}, \bibinfo {author} {\bibfnamefont {P.~B.}\ \bibnamefont {Klar}},
  \bibinfo {author} {\bibfnamefont {H.}~\bibnamefont {Chintakindi}}, \bibinfo
  {author} {\bibfnamefont {A.}~\bibnamefont {Suresh}}, \ and\ \bibinfo {author}
  {\bibfnamefont {L.}~\bibnamefont {Palatinus}},\ }\bibfield  {title} {\enquote
  {\bibinfo {title} {Optimal estimated standard uncertainties of reflection
  intensities for kinematical refinement from {3D} electron diffraction
  data},}\ }\href {\doibase 10.1107/S2053273323005053} {\bibfield  {journal}
  {\bibinfo  {journal} {\textit{Acta Cryst. A}}\ }\textbf {\bibinfo {volume}
  {79}},\ \bibinfo {pages} {427--439} (\bibinfo {year} {2023})}\BibitemShut
  {NoStop}%
\bibitem [{\citenamefont {Bustillo}\ \emph {et~al.}(2021)\citenamefont
  {Bustillo}, \citenamefont {Zeltmann}, \citenamefont {Chen}, \citenamefont
  {Donohue}, \citenamefont {Ciston}, \citenamefont {Ophus},\ and\ \citenamefont
  {Minor}}]{bustillo_4D-STEM_2021}%
  \BibitemOpen
  \bibfield  {author} {\bibinfo {author} {\bibfnamefont {K.~C.}\ \bibnamefont
  {Bustillo}}, \bibinfo {author} {\bibfnamefont {S.~E.}\ \bibnamefont
  {Zeltmann}}, \bibinfo {author} {\bibfnamefont {M.}~\bibnamefont {Chen}},
  \bibinfo {author} {\bibfnamefont {J.}~\bibnamefont {Donohue}}, \bibinfo
  {author} {\bibfnamefont {J.}~\bibnamefont {Ciston}}, \bibinfo {author}
  {\bibfnamefont {C.}~\bibnamefont {Ophus}}, \ and\ \bibinfo {author}
  {\bibfnamefont {A.~M.}\ \bibnamefont {Minor}},\ }\bibfield  {title} {\enquote
  {\bibinfo {title} {{4D-STEM} of beam-sensitive materials},}\ }\href {\doibase
  10.1021/acs.accounts.1c00073} {\bibfield  {journal} {\bibinfo  {journal}
  {\textit{Accounts of Chemical Research}}\ }\textbf {\bibinfo {volume} {54}},\
  \bibinfo {pages} {2543--2551} (\bibinfo {year} {2021})}\BibitemShut {NoStop}%
\bibitem [{\citenamefont {Ophus}(2019)}]{ophus_four-dimensional_2019}%
  \BibitemOpen
  \bibfield  {author} {\bibinfo {author} {\bibfnamefont {C.}~\bibnamefont
  {Ophus}},\ }\bibfield  {title} {\enquote {\bibinfo {title} {Four-dimensional
  scanning transmission electron microscopy ({4D-STEM}): From scanning
  nanodiffraction to ptychography and beyond},}\ }\href {\doibase
  10.1017/S1431927619000497} {\bibfield  {journal} {\bibinfo  {journal}
  {\textit{Microscopy and Microanalysis}}\ }\textbf {\bibinfo {volume} {25}},\
  \bibinfo {pages} {563--582} (\bibinfo {year} {2019})}\BibitemShut {NoStop}%
\bibitem [{\citenamefont {Ercius}\ \emph {et~al.}(2020)\citenamefont {Ercius},
  \citenamefont {Johnson}, \citenamefont {Brown}, \citenamefont {Pelz},
  \citenamefont {Hsu}, \citenamefont {Draney}, \citenamefont {Fong},
  \citenamefont {Goldschmidt}, \citenamefont {Joseph}, \citenamefont {Lee},
  \citenamefont {Ciston}, \citenamefont {Ophus}, \citenamefont {Scott},
  \citenamefont {Selvarajan}, \citenamefont {Paul}, \citenamefont {Skinner},
  \citenamefont {Hanwell}, \citenamefont {Harris}, \citenamefont {Avery},
  \citenamefont {Stezelberger}, \citenamefont {Tindall}, \citenamefont
  {Ramesh}, \citenamefont {Minor},\ and\ \citenamefont
  {Denes}}]{ercius_4d_2020}%
  \BibitemOpen
  \bibfield  {author} {\bibinfo {author} {\bibfnamefont {P.}~\bibnamefont
  {Ercius}}, \bibinfo {author} {\bibfnamefont {I.}~\bibnamefont {Johnson}},
  \bibinfo {author} {\bibfnamefont {H.}~\bibnamefont {Brown}}, \bibinfo
  {author} {\bibfnamefont {P.}~\bibnamefont {Pelz}}, \bibinfo {author}
  {\bibfnamefont {S.-L.}\ \bibnamefont {Hsu}}, \bibinfo {author} {\bibfnamefont
  {B.}~\bibnamefont {Draney}}, \bibinfo {author} {\bibfnamefont
  {E.}~\bibnamefont {Fong}}, \bibinfo {author} {\bibfnamefont {A.}~\bibnamefont
  {Goldschmidt}}, \bibinfo {author} {\bibfnamefont {J.}~\bibnamefont {Joseph}},
  \bibinfo {author} {\bibfnamefont {J.}~\bibnamefont {Lee}}, \bibinfo {author}
  {\bibfnamefont {J.}~\bibnamefont {Ciston}}, \bibinfo {author} {\bibfnamefont
  {C.}~\bibnamefont {Ophus}}, \bibinfo {author} {\bibfnamefont
  {M.}~\bibnamefont {Scott}}, \bibinfo {author} {\bibfnamefont
  {A.}~\bibnamefont {Selvarajan}}, \bibinfo {author} {\bibfnamefont
  {D.}~\bibnamefont {Paul}}, \bibinfo {author} {\bibfnamefont {D.}~\bibnamefont
  {Skinner}}, \bibinfo {author} {\bibfnamefont {M.}~\bibnamefont {Hanwell}},
  \bibinfo {author} {\bibfnamefont {C.}~\bibnamefont {Harris}}, \bibinfo
  {author} {\bibfnamefont {P.}~\bibnamefont {Avery}}, \bibinfo {author}
  {\bibfnamefont {T.}~\bibnamefont {Stezelberger}}, \bibinfo {author}
  {\bibfnamefont {C.}~\bibnamefont {Tindall}}, \bibinfo {author} {\bibfnamefont
  {R.}~\bibnamefont {Ramesh}}, \bibinfo {author} {\bibfnamefont
  {A.}~\bibnamefont {Minor}}, \ and\ \bibinfo {author} {\bibfnamefont
  {P.}~\bibnamefont {Denes}},\ }\bibfield  {title} {\enquote {\bibinfo {title}
  {{The 4D Camera – An 87 kHz Frame-rate Detector for Counted 4D-STEM
  Experiments}},}\ }\href {\doibase 10.1017/S1431927620019753} {\bibfield
  {journal} {\bibinfo  {journal} {\textit{Microscopy and Microanalysis}}\
  }\textbf {\bibinfo {volume} {26}},\ \bibinfo {pages} {1896--1897} (\bibinfo
  {year} {2020})}\BibitemShut {NoStop}%
\bibitem [{\citenamefont {Gammer}\ \emph {et~al.}(2015)\citenamefont {Gammer},
  \citenamefont {Burak~Ozdol}, \citenamefont {Liebscher},\ and\ \citenamefont
  {Minor}}]{gammer_diffraction_2015}%
  \BibitemOpen
  \bibfield  {author} {\bibinfo {author} {\bibfnamefont {C.}~\bibnamefont
  {Gammer}}, \bibinfo {author} {\bibfnamefont {V.}~\bibnamefont {Burak~Ozdol}},
  \bibinfo {author} {\bibfnamefont {C.~H.}\ \bibnamefont {Liebscher}}, \ and\
  \bibinfo {author} {\bibfnamefont {A.~M.}\ \bibnamefont {Minor}},\ }\bibfield
  {title} {\enquote {\bibinfo {title} {Diffraction contrast imaging using
  virtual apertures},}\ }\href {\doibase 10.1016/j.ultramic.2015.03.015}
  {\bibfield  {journal} {\bibinfo  {journal} {\textit{Ultramicroscopy}}\
  }\textbf {\bibinfo {volume} {155}},\ \bibinfo {pages} {1--10} (\bibinfo
  {year} {2015})}\BibitemShut {NoStop}%
\bibitem [{\citenamefont {Bruhn}\ \emph {et~al.}(2021)\citenamefont {Bruhn},
  \citenamefont {Scapin}, \citenamefont {Cheng}, \citenamefont {Mercado},
  \citenamefont {Waterman}, \citenamefont {Ganesh}, \citenamefont {Dallakyan},
  \citenamefont {Read}, \citenamefont {Nieusma}, \citenamefont {Lucier},
  \citenamefont {Mayer}, \citenamefont {Chiang}, \citenamefont {Poweleit},
  \citenamefont {McGilvray}, \citenamefont {Wilson}, \citenamefont {Mashore},
  \citenamefont {Hennessy}, \citenamefont {Thomson}, \citenamefont {Wang},
  \citenamefont {Potter},\ and\ \citenamefont {Carragher}}]{bruhn2023}%
  \BibitemOpen
  \bibfield  {author} {\bibinfo {author} {\bibfnamefont {J.~F.}\ \bibnamefont
  {Bruhn}}, \bibinfo {author} {\bibfnamefont {G.}~\bibnamefont {Scapin}},
  \bibinfo {author} {\bibfnamefont {A.}~\bibnamefont {Cheng}}, \bibinfo
  {author} {\bibfnamefont {B.~Q.}\ \bibnamefont {Mercado}}, \bibinfo {author}
  {\bibfnamefont {D.~G.}\ \bibnamefont {Waterman}}, \bibinfo {author}
  {\bibfnamefont {T.}~\bibnamefont {Ganesh}}, \bibinfo {author} {\bibfnamefont
  {S.}~\bibnamefont {Dallakyan}}, \bibinfo {author} {\bibfnamefont {B.~N.}\
  \bibnamefont {Read}}, \bibinfo {author} {\bibfnamefont {T.}~\bibnamefont
  {Nieusma}}, \bibinfo {author} {\bibfnamefont {K.~W.}\ \bibnamefont {Lucier}},
  \bibinfo {author} {\bibfnamefont {M.~L.}\ \bibnamefont {Mayer}}, \bibinfo
  {author} {\bibfnamefont {N.~J.}\ \bibnamefont {Chiang}}, \bibinfo {author}
  {\bibfnamefont {N.}~\bibnamefont {Poweleit}}, \bibinfo {author}
  {\bibfnamefont {P.~T.}\ \bibnamefont {McGilvray}}, \bibinfo {author}
  {\bibfnamefont {T.~S.}\ \bibnamefont {Wilson}}, \bibinfo {author}
  {\bibfnamefont {M.}~\bibnamefont {Mashore}}, \bibinfo {author} {\bibfnamefont
  {C.}~\bibnamefont {Hennessy}}, \bibinfo {author} {\bibfnamefont
  {S.}~\bibnamefont {Thomson}}, \bibinfo {author} {\bibfnamefont
  {B.}~\bibnamefont {Wang}}, \bibinfo {author} {\bibfnamefont {C.~S.}\
  \bibnamefont {Potter}}, \ and\ \bibinfo {author} {\bibfnamefont
  {B.}~\bibnamefont {Carragher}},\ }\bibfield  {title} {\enquote {\bibinfo
  {title} {Small molecule microcrystal electron diffraction for the
  pharmaceutical industry–lessons learned from examining over fifty
  samples},}\ }\href
  {https://www.frontiersin.org/articles/10.3389/fmolb.2021.648603} {\bibfield
  {journal} {\bibinfo  {journal} {\textit{Frontiers in Molecular Biosciences}}\
  }\textbf {\bibinfo {volume} {8}} (\bibinfo {year} {2021})}\BibitemShut
  {NoStop}%
\bibitem [{\citenamefont {Boggon}\ \emph {et~al.}(2000)\citenamefont {Boggon},
  \citenamefont {Helliwell}, \citenamefont {Judge}, \citenamefont {Olczak},
  \citenamefont {Siddons}, \citenamefont {Snell},\ and\ \citenamefont
  {Stojanoff}}]{boggon_synchrotron_2000}%
  \BibitemOpen
  \bibfield  {author} {\bibinfo {author} {\bibfnamefont {T.~J.}\ \bibnamefont
  {Boggon}}, \bibinfo {author} {\bibfnamefont {J.~R.}\ \bibnamefont
  {Helliwell}}, \bibinfo {author} {\bibfnamefont {R.~A.}\ \bibnamefont
  {Judge}}, \bibinfo {author} {\bibfnamefont {A.}~\bibnamefont {Olczak}},
  \bibinfo {author} {\bibfnamefont {D.~P.}\ \bibnamefont {Siddons}}, \bibinfo
  {author} {\bibfnamefont {E.~H.}\ \bibnamefont {Snell}}, \ and\ \bibinfo
  {author} {\bibfnamefont {V.}~\bibnamefont {Stojanoff}},\ }\bibfield  {title}
  {\enquote {\bibinfo {title} {Synchrotron x-ray reciprocal-space mapping,
  topography and diffraction resolution studies of macromolecular crystal
  quality},}\ }\href {\doibase 10.1107/S0907444900005837} {\bibfield  {journal}
  {\bibinfo  {journal} {\textit{Acta Cryst. D}}\ }\textbf {\bibinfo {volume}
  {56}},\ \bibinfo {pages} {868--880} (\bibinfo {year} {2000})}\BibitemShut
  {NoStop}%
\bibitem [{\citenamefont {Gallagher-Jones}\ \emph {et~al.}(2019)\citenamefont
  {Gallagher-Jones}, \citenamefont {Ophus}, \citenamefont {Bustillo},
  \citenamefont {Boyer}, \citenamefont {Panova}, \citenamefont {Glynn},
  \citenamefont {Zee}, \citenamefont {Ciston}, \citenamefont {Mancia},
  \citenamefont {Minor},\ and\ \citenamefont
  {Rodriguez}}]{gallagher-jones_nanoscale_2019}%
  \BibitemOpen
  \bibfield  {author} {\bibinfo {author} {\bibfnamefont {M.}~\bibnamefont
  {Gallagher-Jones}}, \bibinfo {author} {\bibfnamefont {C.}~\bibnamefont
  {Ophus}}, \bibinfo {author} {\bibfnamefont {K.~C.}\ \bibnamefont {Bustillo}},
  \bibinfo {author} {\bibfnamefont {D.~R.}\ \bibnamefont {Boyer}}, \bibinfo
  {author} {\bibfnamefont {O.}~\bibnamefont {Panova}}, \bibinfo {author}
  {\bibfnamefont {C.}~\bibnamefont {Glynn}}, \bibinfo {author} {\bibfnamefont
  {C.-T.}\ \bibnamefont {Zee}}, \bibinfo {author} {\bibfnamefont
  {J.}~\bibnamefont {Ciston}}, \bibinfo {author} {\bibfnamefont {K.~C.}\
  \bibnamefont {Mancia}}, \bibinfo {author} {\bibfnamefont {A.~M.}\
  \bibnamefont {Minor}}, \ and\ \bibinfo {author} {\bibfnamefont {J.~A.}\
  \bibnamefont {Rodriguez}},\ }\bibfield  {title} {\enquote {\bibinfo {title}
  {Nanoscale mosaicity revealed in peptide microcrystals by scanning electron
  nanodiffraction},}\ }\href {\doibase 10.1038/s42003-018-0263-8} {\bibfield
  {journal} {\bibinfo  {journal} {\textit{Communications Biology}}\ }\textbf
  {\bibinfo {volume} {2}},\ \bibinfo {pages} {1--8} (\bibinfo {year}
  {2019})}\BibitemShut {NoStop}%
\bibitem [{\citenamefont {Jones}\ \emph {et~al.}(2019)\citenamefont {Jones},
  \citenamefont {Asay}, \citenamefont {Kim}, \citenamefont {Kleinsasser},
  \citenamefont {Saha}, \citenamefont {Fulton}, \citenamefont {Berkley},
  \citenamefont {Cascio}, \citenamefont {Malyutin}, \citenamefont {Conley},
  \citenamefont {Stoltz}, \citenamefont {Lavallo}, \citenamefont {Rodríguez},\
  and\ \citenamefont {Nelson}}]{jones_organomet}%
  \BibitemOpen
  \bibfield  {author} {\bibinfo {author} {\bibfnamefont {C.~G.}\ \bibnamefont
  {Jones}}, \bibinfo {author} {\bibfnamefont {M.}~\bibnamefont {Asay}},
  \bibinfo {author} {\bibfnamefont {L.~J.}\ \bibnamefont {Kim}}, \bibinfo
  {author} {\bibfnamefont {J.~F.}\ \bibnamefont {Kleinsasser}}, \bibinfo
  {author} {\bibfnamefont {A.}~\bibnamefont {Saha}}, \bibinfo {author}
  {\bibfnamefont {T.~J.}\ \bibnamefont {Fulton}}, \bibinfo {author}
  {\bibfnamefont {K.~R.}\ \bibnamefont {Berkley}}, \bibinfo {author}
  {\bibfnamefont {D.}~\bibnamefont {Cascio}}, \bibinfo {author} {\bibfnamefont
  {A.~G.}\ \bibnamefont {Malyutin}}, \bibinfo {author} {\bibfnamefont {M.~P.}\
  \bibnamefont {Conley}}, \bibinfo {author} {\bibfnamefont {B.~M.}\
  \bibnamefont {Stoltz}}, \bibinfo {author} {\bibfnamefont {V.}~\bibnamefont
  {Lavallo}}, \bibinfo {author} {\bibfnamefont {J.~A.}\ \bibnamefont
  {Rodríguez}}, \ and\ \bibinfo {author} {\bibfnamefont {H.~M.}\ \bibnamefont
  {Nelson}},\ }\bibfield  {title} {\enquote {\bibinfo {title} {Characterization
  of reactive organometallic species via {microED}},}\ }\href {\doibase
  10.1021/acscentsci.9b00403} {\bibfield  {journal} {\bibinfo  {journal}
  {\textit{ACS Central Science}}\ }\textbf {\bibinfo {volume} {5}},\ \bibinfo
  {pages} {1507--1513} (\bibinfo {year} {2019})}\BibitemShut {NoStop}%
\bibitem [{\citenamefont {Alexander}\ and\ \citenamefont
  {Charlesby}(1954)}]{alexander_energy_1954}%
  \BibitemOpen
  \bibfield  {author} {\bibinfo {author} {\bibfnamefont {P.}~\bibnamefont
  {Alexander}}\ and\ \bibinfo {author} {\bibfnamefont {A.}~\bibnamefont
  {Charlesby}},\ }\bibfield  {title} {\enquote {\bibinfo {title} {Energy
  transfer in macromolecules exposed to ionizing radiations},}\ }\href
  {\doibase 10.1038/173578a0} {\bibfield  {journal} {\bibinfo  {journal}
  {\textit{Nature}}\ }\textbf {\bibinfo {volume} {173}},\ \bibinfo {pages}
  {578--579} (\bibinfo {year} {1954})}\BibitemShut {NoStop}%
\bibitem [{\citenamefont {Liming}\ and\ \citenamefont
  {Gordy}(1968)}]{liming_hydrogen-addition_1968}%
  \BibitemOpen
  \bibfield  {author} {\bibinfo {author} {\bibfnamefont {F.~G.}\ \bibnamefont
  {Liming}}\ and\ \bibinfo {author} {\bibfnamefont {W.}~\bibnamefont {Gordy}},\
  }\bibfield  {title} {\enquote {\bibinfo {title} {Hydrogen-addition radicals
  formed in the aromatic rings of amino acids, polyamino acids, and
  proteins.}}\ }\href {\doibase 10.1073/pnas.60.3.794} {\bibfield  {journal}
  {\bibinfo  {journal} {\textit{Proceedings of the National Academy of
  Sciences}}\ }\textbf {\bibinfo {volume} {60}},\ \bibinfo {pages} {794--801}
  (\bibinfo {year} {1968})}\BibitemShut {NoStop}%
\bibitem [{\citenamefont {Lin}(1974)}]{lin_electron_1974}%
  \BibitemOpen
  \bibfield  {author} {\bibinfo {author} {\bibfnamefont {S.~D.}\ \bibnamefont
  {Lin}},\ }\bibfield  {title} {\enquote {\bibinfo {title} {Electron radiation
  damage of thin films of glycine, diglycine, and aromatic amino acids},}\
  }\href {\doibase 10.2307/3574071} {\bibfield  {journal} {\bibinfo  {journal}
  {\textit{Radiation Research}}\ }\textbf {\bibinfo {volume} {59}},\ \bibinfo
  {pages} {521--536} (\bibinfo {year} {1974})}\BibitemShut {NoStop}%
\bibitem [{\citenamefont {Howie}, \citenamefont {Rocca},\ and\ \citenamefont
  {Valdrè}(1985)}]{howie_electron_1985}%
  \BibitemOpen
  \bibfield  {author} {\bibinfo {author} {\bibfnamefont {A.}~\bibnamefont
  {Howie}}, \bibinfo {author} {\bibfnamefont {F.~J.}\ \bibnamefont {Rocca}}, \
  and\ \bibinfo {author} {\bibfnamefont {U.}~\bibnamefont {Valdrè}},\
  }\bibfield  {title} {\enquote {\bibinfo {title} {Electron beam ionization
  damage processes in p-terphenyl},}\ }\href {\doibase
  10.1080/13642818508240634} {\bibfield  {journal} {\bibinfo  {journal}
  {\textit{Philosophical Magazine B}}\ }\textbf {\bibinfo {volume} {52}},\
  \bibinfo {pages} {751--757} (\bibinfo {year} {1985})}\BibitemShut {NoStop}%
\bibitem [{\citenamefont {Schulze-Briese}\ \emph {et~al.}(2005)\citenamefont
  {Schulze-Briese}, \citenamefont {Wagner}, \citenamefont {Tomizaki},\ and\
  \citenamefont {Oetiker}}]{Schulze-Briese:xn0007}%
  \BibitemOpen
  \bibfield  {author} {\bibinfo {author} {\bibfnamefont {C.}~\bibnamefont
  {Schulze-Briese}}, \bibinfo {author} {\bibfnamefont {A.}~\bibnamefont
  {Wagner}}, \bibinfo {author} {\bibfnamefont {T.}~\bibnamefont {Tomizaki}}, \
  and\ \bibinfo {author} {\bibfnamefont {M.}~\bibnamefont {Oetiker}},\
  }\bibfield  {title} {\enquote {\bibinfo {title} {{Beam-size effects in
  radiation damage in insulin and thaumatin crystals}},}\ }\href {\doibase
  10.1107/S0909049505003298} {\bibfield  {journal} {\bibinfo  {journal}
  {\textit{Journal of Synchrotron Radiation}}\ }\textbf {\bibinfo {volume}
  {12}},\ \bibinfo {pages} {261--267} (\bibinfo {year} {2005})}\BibitemShut
  {NoStop}%
\bibitem [{\citenamefont {Diederichs}\ and\ \citenamefont
  {Karplus}(1997)}]{diederichs_improved_1997}%
  \BibitemOpen
  \bibfield  {author} {\bibinfo {author} {\bibfnamefont {K.}~\bibnamefont
  {Diederichs}}\ and\ \bibinfo {author} {\bibfnamefont {P.~A.}\ \bibnamefont
  {Karplus}},\ }\bibfield  {title} {\enquote {\bibinfo {title} {{Improved
  R-factors for diffraction data analysis in macromolecular
  crystallography}},}\ }\href {\doibase 10.1038/nsb0497-269} {\bibfield
  {journal} {\bibinfo  {journal} {\textit{Nature Structural Biology}}\ }\textbf
  {\bibinfo {volume} {4}},\ \bibinfo {pages} {269--275} (\bibinfo {year}
  {1997})}\BibitemShut {NoStop}%
\bibitem [{\citenamefont {Owen}\ \emph {et~al.}(2014)\citenamefont {Owen},
  \citenamefont {Paterson}, \citenamefont {Axford}, \citenamefont {Aishima},
  \citenamefont {Schulze-Briese}, \citenamefont {Ren}, \citenamefont {Fry},
  \citenamefont {Stuart},\ and\ \citenamefont {Evans}}]{owen_exploiting_2014}%
  \BibitemOpen
  \bibfield  {author} {\bibinfo {author} {\bibfnamefont {R.~L.}\ \bibnamefont
  {Owen}}, \bibinfo {author} {\bibfnamefont {N.}~\bibnamefont {Paterson}},
  \bibinfo {author} {\bibfnamefont {D.}~\bibnamefont {Axford}}, \bibinfo
  {author} {\bibfnamefont {J.}~\bibnamefont {Aishima}}, \bibinfo {author}
  {\bibfnamefont {C.}~\bibnamefont {Schulze-Briese}}, \bibinfo {author}
  {\bibfnamefont {J.}~\bibnamefont {Ren}}, \bibinfo {author} {\bibfnamefont
  {E.~E.}\ \bibnamefont {Fry}}, \bibinfo {author} {\bibfnamefont {D.~I.}\
  \bibnamefont {Stuart}}, \ and\ \bibinfo {author} {\bibfnamefont
  {G.}~\bibnamefont {Evans}},\ }\bibfield  {title} {\enquote {\bibinfo {title}
  {Exploiting fast detectors to enter a new dimension in room-temperature
  crystallography},}\ }\href {\doibase 10.1107/S1399004714005379} {\bibfield
  {journal} {\bibinfo  {journal} {\textit{Acta Cryst. D}}\ }\textbf {\bibinfo
  {volume} {70}},\ \bibinfo {pages} {1248--1256} (\bibinfo {year}
  {2014})}\BibitemShut {NoStop}%
\bibitem [{\citenamefont {O'Hagan}(2008)}]{ohagan_understanding_2008}%
  \BibitemOpen
  \bibfield  {author} {\bibinfo {author} {\bibfnamefont {D.}~\bibnamefont
  {O'Hagan}},\ }\bibfield  {title} {\enquote {\bibinfo {title} {{Understanding
  organofluorine chemistry. An introduction to the C–F bond}},}\ }\href
  {\doibase 10.1039/B711844A} {\bibfield  {journal} {\bibinfo  {journal}
  {\textit{Chem. Soc. Rev.}}\ }\textbf {\bibinfo {volume} {37}},\ \bibinfo
  {pages} {308--319} (\bibinfo {year} {2008})}\BibitemShut {NoStop}%
\bibitem [{\citenamefont {Hachtel}\ \emph {et~al.}(2019)\citenamefont
  {Hachtel}, \citenamefont {Huang}, \citenamefont {Popovs}, \citenamefont
  {Jansone-Popova}, \citenamefont {Keum}, \citenamefont {Jakowski},
  \citenamefont {Lovejoy}, \citenamefont {Dellby}, \citenamefont {Krivanek},\
  and\ \citenamefont {Idrobo}}]{hachtel_identification_2019}%
  \BibitemOpen
  \bibfield  {author} {\bibinfo {author} {\bibfnamefont {J.~A.}\ \bibnamefont
  {Hachtel}}, \bibinfo {author} {\bibfnamefont {J.}~\bibnamefont {Huang}},
  \bibinfo {author} {\bibfnamefont {I.}~\bibnamefont {Popovs}}, \bibinfo
  {author} {\bibfnamefont {S.}~\bibnamefont {Jansone-Popova}}, \bibinfo
  {author} {\bibfnamefont {J.~K.}\ \bibnamefont {Keum}}, \bibinfo {author}
  {\bibfnamefont {J.}~\bibnamefont {Jakowski}}, \bibinfo {author}
  {\bibfnamefont {T.~C.}\ \bibnamefont {Lovejoy}}, \bibinfo {author}
  {\bibfnamefont {N.}~\bibnamefont {Dellby}}, \bibinfo {author} {\bibfnamefont
  {O.~L.}\ \bibnamefont {Krivanek}}, \ and\ \bibinfo {author} {\bibfnamefont
  {J.~C.}\ \bibnamefont {Idrobo}},\ }\bibfield  {title} {\enquote {\bibinfo
  {title} {Identification of site-specific isotopic labels by vibrational
  spectroscopy in the electron microscope},}\ }\href {\doibase
  10.1126/science.aav5845} {\bibfield  {journal} {\bibinfo  {journal}
  {\textit{Science}}\ }\textbf {\bibinfo {volume} {363}},\ \bibinfo {pages}
  {525--528} (\bibinfo {year} {2019})}\BibitemShut {NoStop}%
\bibitem [{\citenamefont {Warkentin}\ \emph {et~al.}(2013)\citenamefont
  {Warkentin}, \citenamefont {Hopkins}, \citenamefont {Badeau}, \citenamefont
  {Mulichak}, \citenamefont {Keefe},\ and\ \citenamefont
  {Thorne}}]{warkentin_global_2013}%
  \BibitemOpen
  \bibfield  {author} {\bibinfo {author} {\bibfnamefont {M.}~\bibnamefont
  {Warkentin}}, \bibinfo {author} {\bibfnamefont {J.~B.}\ \bibnamefont
  {Hopkins}}, \bibinfo {author} {\bibfnamefont {R.}~\bibnamefont {Badeau}},
  \bibinfo {author} {\bibfnamefont {A.~M.}\ \bibnamefont {Mulichak}}, \bibinfo
  {author} {\bibfnamefont {L.~J.}\ \bibnamefont {Keefe}}, \ and\ \bibinfo
  {author} {\bibfnamefont {R.~E.}\ \bibnamefont {Thorne}},\ }\bibfield  {title}
  {\enquote {\bibinfo {title} {Global radiation damage: temperature dependence,
  time dependence and how to outrun it},}\ }\href {\doibase
  10.1107/S0909049512048303} {\bibfield  {journal} {\bibinfo  {journal}
  {\textit{Journal of Synchrotron Radiation}}\ }\textbf {\bibinfo {volume}
  {20}},\ \bibinfo {pages} {7--13} (\bibinfo {year} {2013})}\BibitemShut
  {NoStop}%
\bibitem [{\citenamefont {Warkentin}\ \emph {et~al.}(2017)\citenamefont
  {Warkentin}, \citenamefont {Atakisi}, \citenamefont {Hopkins}, \citenamefont
  {Walko},\ and\ \citenamefont {Thorne}}]{warkentin_lifetimes_2017}%
  \BibitemOpen
  \bibfield  {author} {\bibinfo {author} {\bibfnamefont {M.~A.}\ \bibnamefont
  {Warkentin}}, \bibinfo {author} {\bibfnamefont {H.}~\bibnamefont {Atakisi}},
  \bibinfo {author} {\bibfnamefont {J.~B.}\ \bibnamefont {Hopkins}}, \bibinfo
  {author} {\bibfnamefont {D.}~\bibnamefont {Walko}}, \ and\ \bibinfo {author}
  {\bibfnamefont {R.~E.}\ \bibnamefont {Thorne}},\ }\bibfield  {title}
  {\enquote {\bibinfo {title} {Lifetimes and spatio-temporal response of
  protein crystals in intense x-ray microbeams},}\ }\href {\doibase
  10.1107/S2052252517013495} {\bibfield  {journal} {\bibinfo  {journal}
  {{\textit{IUCrJ}}}\ }\textbf {\bibinfo {volume} {4}},\ \bibinfo {pages}
  {785--794} (\bibinfo {year} {2017})}\BibitemShut {NoStop}%
\bibitem [{\citenamefont {Bammes}\ \emph {et~al.}(2010)\citenamefont {Bammes},
  \citenamefont {Jakana}, \citenamefont {Schmid},\ and\ \citenamefont
  {Chiu}}]{bammes_radiation_2010}%
  \BibitemOpen
  \bibfield  {author} {\bibinfo {author} {\bibfnamefont {B.~E.}\ \bibnamefont
  {Bammes}}, \bibinfo {author} {\bibfnamefont {J.}~\bibnamefont {Jakana}},
  \bibinfo {author} {\bibfnamefont {M.~F.}\ \bibnamefont {Schmid}}, \ and\
  \bibinfo {author} {\bibfnamefont {W.}~\bibnamefont {Chiu}},\ }\bibfield
  {title} {\enquote {\bibinfo {title} {{Radiation damage effects at four
  specimen temperatures from 4 to 100 K}},}\ }\href {\doibase
  10.1016/j.jsb.2009.11.001} {\bibfield  {journal} {\bibinfo  {journal}
  {\textit{Journal of Structural Biology}}\ }\textbf {\bibinfo {volume}
  {169}},\ \bibinfo {pages} {331--341} (\bibinfo {year} {2010})}\BibitemShut
  {NoStop}%
\end{thebibliography}%

\end{document}